\documentclass[fleqn,11pt,thmsb,draft]{article}
\usepackage{graphicx}
\usepackage{amsmath}
\usepackage{amsfonts}
\usepackage{amssymb}
\newtheorem{theorem}{Theorem}[section]

\newtheorem{definition}[theorem]{Definition}
\newtheorem{example}[theorem]{Example}

\newtheorem{problem}[theorem]{Problem}
\newtheorem{proposition}[theorem]{Proposition}

\newenvironment{proof}[1][Proof]{\noindent\textbf{#1.} }{}

\begin{document}

\title{L-Fuzzy Valued Inclusion Measure, \\L-Fuzzy Similarity and L-Fuzzy Distance}
\author{Ath. Kehagias and M. Konstantinidou\\Box 464, Division of Mathematics\\Dept. of Mathematics, Physical and Computational Sciences, \\Faculty of Engineering, Aristotle University of Thessaloniki \\Thessaloniki, GR\ 54006, GREECE\\email:\ kehagias@egnatia.ee.auth.gr}
\date{20 July 2001}
\maketitle
\begin{abstract}
The starting point of this paper is the introduction of a new \emph{measure of
inclusion} of fuzzy set $A$ in fuzzy set $B$. Previously used inclusion
measures take values in the interval [0,1]; the inclusion measure proposed
here takes values in a \emph{Boolean lattice}. In other words, inclusion is
viewed as an \emph{L-fuzzy valued relation} between fuzzy sets. This relation
is reflexive, antisymmetric and transitive, i.e. it is a \emph{fuzzy order
relation}; in addition it possesess a number of properties which various
authors have postulated as axiomatically appropriate for an inclusion measure.
We also define an L-fuzzy valued measure of \emph{similarity} between fuzzy
sets and and an L-fuzzy valued \emph{distance} function between fuzzy sets;
these possess properties analogous to the ones of real-valued similarity and
distance functions.

\noindent\textbf{Keywords:} Fuzzy Relations, inclusion measure, subsethood,
L-fuzzy sets, similarity, distance, transitivity.
\end{abstract}

\section{Introduction}

\label{sec01}

A \emph{measure of inclusion }(also called a \emph{subsethood measure}%
)\emph{\ }is a \emph{relation }between fuzzy sets $A$ and $B$, which indicates
the degree to which $A$ is contained in (is a subset of) $B$. Many measures of
inclusion have been proposed in the literature; these usually take values
either in $\{0,1\}$ or in [0,1]. The first case corresponds to a \emph{crisp}
relation:\ $A$ is either contained or not contained in $B$, while the second
case corrresponds to a \emph{fuzzy} relation:\ $A$ is contained in $B$ to a
certain degree. A related concept is that of a \emph{measure of similarity
}between fuzzy sets; a \emph{measure of similarity }is a relation which can be
seen as a fuzzification of a crisp \emph{equivalence relation}.

In this paper we introduce an inclusion measure, denoted by $I(A,B)$, i.e.
$I(A,B)$ denotes the degree to which fuzzy set $A$ is included in fuzzy set $B
$. We also introduce a related similarity measure denoted by $S(A,B)$ and a
\emph{distance between fuzzy sets }$D(A,B)$. The main difference between our
work and that of other authors is the following:\ traditionally, inclusion
measures, similarities and distances take values in [0,1] or some other
\emph{totally ordered }subset of the \emph{real }interval [0,$\infty$); in
this paper, on the other hand, the range of $I(.,.)$, $S(.,.)$ and $D(.,.)$ is
a \emph{Boolean lattice} $\mathbf{B}$. Since $\mathbf{B}$ is a
\emph{partially} (but not totally) ordered set, it follows that the proposed
inclusion, similarity and distance are \emph{L-fuzzy valued relations between
fuzzy sets}. For the sake of brevity we will occasionally refer to quantities
which take values in a totally ordered set as \emph{``scalar''} and to
quantities which take take values in a partially ordered set as
\emph{``vector''}. The rationale for this terminology will become obvious in
the sequel.

The proposed inclusion measure $I(A,B)$ is reflexive, symmetric and
\emph{transitive }-- hence is a \emph{L-fuzzy order relation}. In addition,
$I(A,B) $ possesess a number of attractive properties which various authors
have postulated as axiomatically appropriate for an inclusion measure. Similar
remarks hold for the proposed similarity $S(A,B)$ and distance $D(A,B)$.

Let us now briefly review related previous work. Zadeh, in the seminal paper
\cite{Zadeh01}, gives the first definition of fuzzy set inclusion; in Zadeh's
formulation, inclusion is a \emph{crisp} relation (i.e. a fuzzy set $A$ is
either included or not included in a fuzzy set $B$). Also, there is a popular
concept of fuzzy set inclusion which is essentially a fuzzy analog of
conditional probability (in this paper it appears as inclusion measure no.1 in
Section \ref{sec0302}). This appears in many works (see for instance
\cite{Bodj01}). An interesting variation of this measure (in this paper it
appears as inclusion measure no.2 in Section \ref{sec0302}) appears in
\cite{JFan01,Willmott1,Vgk01,Vgk04,Vgk03}.

Several authors take an \emph{axiomatic} approach to the study of inclusion
measure, i.e. they provide a list of properties (``axioms'') which a
``reasonable'' inclusion measure must satisfy and then they examine whether a
particular inclusion measure or a family of inclusion measures possesses these
properties. A prime example of this approach is Sinha and Dougherty's
\cite{Sinh01}. They list nine properties that a ``reasonable'' inclusion
measure should have and then proceed to introduce inclusion measures which
have these properties.

A somewhat different point of view is taken by Bandler and Kohout in
\cite{Band01}. These authors obtain several inclusion measures from
\emph{fuzzy implication operators}. A related approach is that of
\cite{Willmott1,Willmott2} where the transitivity of inclusion measure is also
studied. The issue of transitivity is also studied in \cite{Kundu1}, where a
\ $\wedge$--transitive inclusion measure is introduced.

Axiomatization and connection to implication operators are combined in a
number of papers. Young combines the above approaches in \cite{Virg01} and
also connects inclusion measure to fuzzy entropy. Young gives various examples
of inclusion measures listed in the literature and examines which of the
properties she proposes are satisfied by these inclusion measures. She also
gives the connection to Bandler and Kohout's work on fuzzy implication
operators. Finally, Young presents an application of inclusion measure to
tuning fuzzy logic rules. In \cite{Bur01}, Burillo et al. introduce a
particular family of fuzzy implication operators (a form of generalized
Lukasiewicz operators) and show that the inclusion measures obtained from this
family (in Bandler and Kohout's manner), satisfy Sinha and Dougherty's axioms
for inclusion measures. In \cite{JFan01} J. Fan et al. discuss the connections
between inclusion measure, fuzzy entropy and fuzzy implication. They comment
on Young's axioms and propose their own list of axioms. They also give some
conditions to check if a function is an inclusion measure. Finally, they
present an application of inclusion measure to clustering validity.

All the inclusion measures discussed up to this point take values in a totally
ordered set. In \cite{Bustinc01}, Bustince introduces an inclusion measure
which takes values in the partially ordered set of interval valued fuzzy sets.
Bustince also relates inclusion to implication operators.

A somewhat different view on inclusion measures is taken in \cite{Bouch01}.
The authors classify measures (of similarity, satisfiability, inclusion,
dissimilarity etc.) using a very general scheme to generate such measures from
elementary functions; the corresponding properties are also quite general. The
authors also consider the \emph{aggreggation} of several local measures of
similarity into a global one.

In some of the above mentioned works, the connection between inclusion and
similarity measures is discussed. Fuzzy similarity has received much more
attention than fuzzy inclusion and the corresponding literature is very
extensive; here we discuss a small number of papers, which have a point of
view similar to ours.

Pappis and his collaborators have issued a series of papers
\cite{Pappis01,Pappis02,Pappis03} which take an axiomatic view to similarity
measures. In \cite{JFan02} J. Fan gives axioms for entropy, distance and
similarity. Liu in \cite{Liu1} proposes an alternative set of axioms. X. Wang
et al. criticize Pappis' work in \cite{Wang02}, present a modified definition
of similarity and explore its connection to Bandler and Kohout's work. W. Wang
in \cite{Wang01} adopts Liu's similarity axioms and introduces two new
similarity measures. Ovchinnikov in \cite{Ovchi01} discusses L-fuzzy
relations, i.e. relations taking values in lattices and mentions similarity
relations in this context.

Returning to the topic of inclusion measures, we should mention
\cite{Vgk01,Vgk04,Vgk03} which take a somewhat different view towards
inclusion. Namely, the authors introduce a fuzzy measure of the inclusion of a
\emph{crisp} set $A$ in a \emph{crisp} set $B$ and develop an extensive
methodology which uses their inclusion measure for clustering and
classification applications.

As we have already pointed out, practically all of the work on fuzzy inclusion
and similarity uses measures taking values in the \emph{totally ordered
}interval [0,1]. A notable exception is \cite{Bustinc01} (which uses interval
valued similarity measures). A more general discussion of L-fuzzy valued
relations can be found in, for example, \cite{Fodor01} and \cite{Seselj01}.

\section{Preliminaries}

\label{sec02}

\subsection{Fuzzy and Crisp Sets}

\label{sec0201}All subsequent discussion makes use of two fundamental sets.
First, we have a \emph{universe of discourse }denoted by $U$; this can be
quite general, i.e. no special structure or properties are assumed (in
particular, $U$ can be finite, countable or uncountable and we assume no order
on the elements of $U$). Second, we have a \emph{totally ordered} set denoted
by $L$; in this paper we take $L=[0,1]\ $(but most of our results remain valid
in case $L$ is a finite set $\{a_{1},a_{2},...,a_{2N+1}\}$ with $0=a_{1}$
$<$%
\ $a_{2}$
$<$%
\ ...
$<$%
$a_{N+1}=\frac{1}{2}$
$<$%
...
$<$%
$a_{2N}$
$<$%
\ $a_{2N+1}=1$). Elements of both $U$ and $L$ will be denoted by lowercase
letters. For elements $x,y\in L$ we will use $x\leq y$ to denote that $x$ is
less than or equal to $y$; the symbols $<,$ $\geq,$ $>$ also have their usual
significance. The minimum of $x$ and $y$ will be denoted by $x\wedge y$ and
the maximum of $x$ and $y$ will be denoted by $x\vee y$. We also introduce
negation of the elements of $L$ (denoted by $^{\prime}$) which is defined by:
$a^{\prime}\doteq1-a$; here ``$-$'' indicates the usual subtraction of real
numbers. Hence $(L,\leq,\wedge,\vee,^{\prime})$ is a \emph{de Morgan lattice},
i.e. a bounded lattice, \emph{distributive }with respect to the $\wedge,\vee$
operations, \emph{order-inverting }with respect to the $^{\prime}$ operation
and satisfying de Morgan's laws \cite{Ngu01}. In addition, $(L,\leq)$ is a
complete lattice and totally ordered.

We will be concerned with fuzzy (sub)sets of $U$; fuzzy sets are identified
with their membership functions, which take values in $L$. In other words, a
fuzzy set is a function $A:U\rightarrow L$. Fuzzy sets will be denoted by
uppercase letters: $A,B,C$, ... , with two exceptions:\ the empty fuzzy set
will be denoted by \underline{$0$} (this is equivalent to $\emptyset$) and the
universal fuzzy set will be denoted by \underline{1} (this is equivalent to
$U$). In case $U$ is a denumerable set, then we can represent a fuzzy set $A$
in vector notation $A=[A_{1},A_{2},...]$. Hence the degree to which element
$u$ belongs in set $A$ is denoted by $A_{u}$. The family of all fuzzy sets
(the fuzzy \emph{powerset} of $U$) will be denoted by $\mathbf{F}(U)$ or
simply by $\mathbf{F}$.

\begin{example}
\label{exa01}Take $U=\{1,2,3,4\}$. Then we have $\underline{1}=[1,1,1,1]$
(i.e. $\underline{1}_{1}=1,\underline{1}_{2}=1,\underline{1}_{3}%
=1,\underline{1}_{4}=1$) and $\underline{0}=[0,0,0,0]$. Other examples of
fuzzy sets are $A=[0.2,0.3,0.0,0.9]$ (i.e. $A_{1}=0.2,A_{2}=0.3,A_{3}%
=0.0,A_{4}=0.9 $), $B=[0.3,0.4,0.2,0.8]$, $C=[0.3,0.5,0.8,1.0]$.
\end{example}

Crisp (sub)sets of $U$ are (obviously)\ special cases of fuzzy sets and are
also identified with their membership (or characteristic) functions, which
take values in $\{0,1\}$. In other words, a fuzzy set is a function
$\Theta:U\rightarrow\{0,1\}$. To underscore the fact that a set is crisp we
will usually denote it by an uppercase \emph{Greek} letter such as:
$\Theta,\Phi,\Xi$, ... . The family of all crisp sets (the crisp powerset of
$U$) will be denoted by $\mathbf{B}(U)$ or simply by $\mathbf{B}$. Clearly
$\mathbf{B\subseteq F}$.

\begin{example}
\label{exa02}Take $U=\{1,2,3,4\}$. Examples of crisp sets are $\Theta
=[0,1,1,1]$ and $\Phi=[1,1,0,1]$.
\end{example}

\subsection{The $(\mathbf{F},\leq)$ Lattice}

\label{sec0202}

We have already remarked that $(L,\leq,\wedge,\vee,^{\prime})$ is a totally
ordered \emph{de Morgan lattice}. We now define an order on elements of
$\mathbf{F}$. This is obtained from the ``elementwise'' order and is denoted
(without danger of confusion)\ by the same symbol $\leq$. I.e., for all
$A,B\in\mathbf{F}$ we define
\[
A\leq B\text{ }\Leftrightarrow\text{ (for all }u\in U\text{ we have }A_{u}\leq
B_{u}\text{).}%
\]
It is easy to see that $\leq$ is an order on $\mathbf{F}$. More rigorously,
\[
\mathbf{F\doteq}\underset{|U|\text{ times}}{\underbrace{L\times L\times
...\times L}.}%
\]
and $(\mathbf{F,\leq)}$ is the direct product lattice \cite{Birk01}
\[
(\mathbf{F,\leq)\doteq}\underset{|U|\text{ times}}{\underbrace{(L,\leq
)\times(L,\leq)\times...\times(L,\leq)}.}\text{ }%
\]
($\mathbf{F,\leq)}$ $\ $is a \emph{partially} ordered set, i.e. there exist
elements $X,Y\in\mathbf{F}$ such that neither $X\leq Y$ nor $Y\leq X$; this is
denoted by $X||Y$ ($X$ and $Y$ are \emph{incomparable}). Furthermore, as a
product of complete lattices, ($\mathbf{F},\leq$) is complete \cite{Davey1}.

The sup and inf operations on elements of $\mathbf{F}$ (with respect to $\leq
$) can be obtained from the elementwise min and max operations. I.e. for all
$A,B\in\mathbf{F}$ the $\inf(A,B)$ exist; it is denoted (without danger of
confusion) by $A\wedge B$ and defined for all $u\in U$ by
\[
(A\wedge B)_{u}\doteq A_{u}\wedge B_{u}.
\]
(Notice the notation above:\ $A\wedge B$ is a function; the value of the
function at $u$ is denoted by $(A\wedge B)_{u}$.) Similarly, for all
$A,B\in\mathbf{F}$ the $\sup(A,B)$ exists; it is denoted (without danger of
confusion) by $A\vee B$ and defined for all $u\in U$ by
\[
(A\vee B)_{u}\doteq A_{u}\vee B_{u}.
\]
Finally, complementation is defined on $\mathbf{F}$ in terms of elementwise
complementation: for all $A\in\mathbf{F}$ and $u\in U$ we have $(A^{\prime
})_{u}=(A_{u})^{\prime}$.

\begin{example}
\label{exa04}We continue using the sets of the previous examples. It can be
seen that for $u=1,2,3,4$ we have $A_{u}\leq C_{u}$; hence $A\leq C$, $A\vee
C=C$, $A\wedge C=A$. $A$ and $B$ are incomparable: $A||B$. We have $A\vee B$ =
$[0.3,0.4,0.2,0.9]$ and $A\wedge B$ = $[0.2,0.3,0.0,0.8]$; we write (for
instance)\ ($A\wedge B)_{1}$ = $0.2$. Similarly $B\vee\Phi$ = $[1,1,0.2,1]$.
Finally, $A^{\prime}=[0.8,0.7,1.0,0.1]$.
\end{example}

It is easy to prove that $(\mathbf{F},\leq,\wedge,\vee,^{\prime})$ is a de
Morgan lattice. It is also easy to prove that $(\mathbf{B},\leq,\wedge
,\vee,^{\prime})$ is a \emph{Boolean lattice }(or \emph{Boolean algebra}), a
complete lattice and a sublattice of $(\mathbf{F},\leq,\wedge,\vee,^{\prime})$.

\subsection{Fuzzy Relations}

\label{sec0203}

In the sequel we will often use \emph{fuzzy relations }on \emph{subsets of
}$U$ (e.g. inclusion, similarity, distance). In this paper, the fuzzy
relations of interest are functions of the form $R:$ $\mathbf{F}%
\times\mathbf{F}\rightarrow\mathbf{B}$. Adapting well known definitions
\cite{Klir01} to this context we have the following.

\begin{definition}
\label{def0201}A fuzzy relation $R$ is called \emph{reflexive} iff $\forall
A\in\mathbf{F}$ we have $R(A,A)=\underline{1}$.
\end{definition}

\begin{definition}
\label{def0202}A fuzzy relation $R$ is called \emph{symmetric }iff $\forall
A,B\in\mathbf{F}$ we have $R(A,B)=R(B,A)$.
\end{definition}

\begin{definition}
\label{def0203}A fuzzy relation $R$ is called \emph{antisymmetric }iff
$\forall A,B\in\mathbf{F}$ we have
\[
R(A,B)=R(B,A)>\underline{0}\Rightarrow A=B.
\]
\end{definition}

\begin{definition}
\label{def0204}A fuzzy relation $R$ is called $\wedge$-\emph{transitive }iff
$\forall A,B\in\mathbf{F}$ we have
\[
R(A,B)\geq\sup_{C\in\mathbf{F}}[R(A,C)\wedge R(C,B)].
\]
\end{definition}

\begin{definition}
\label{def0205}A fuzzy relation $R$ is called $\vee$-\emph{transitive }iff
$\forall A,B\in\mathbf{F}$ we have
\[
R(A,B)\leq\inf_{C\in\mathbf{F}}[R(A,C)\vee R(C,B)].
\]
\end{definition}

\noindent\textbf{Remark}. In the above definitions the indicated sup and inf
exist always because $(\mathbf{B},\leq,\wedge,\vee,^{\prime})$ is a complete lattice.

\begin{definition}
\label{def0206}A fuzzy relation $R$ is called a \emph{fuzzy similarity
relation} (or \emph{fuzzy equivalence relation}) if it is reflexive, symmetric
and $\wedge$-transitive.
\end{definition}

\begin{definition}
\label{def0207}A fuzzy relation $R$ is called a \emph{fuzzy order relation }if
it is reflexive, antisymmetric and $\wedge$-transitive.
\end{definition}

Note that $R(A,B)$ is a function with \emph{domain }$\mathbf{F}\times
\mathbf{F}$ and \emph{range} $\mathbf{B}$.To denote the value of $R(A,B)$ at
point $u\in U$, we use the notation$\ R_{u}(A,B).$\footnote{It is slightly
unusual to place the argument of a function as a subscript, but this notation
will prove advantageous in the sequel.}

\section{The L-Fuzzy Inclusion Measure}

\label{sec03}

\subsection{Definitions and Properties}

\label{sec0301}

\begin{definition}
\label{def0301}For all $A,B\in\mathbf{F}$, the \emph{measure of inclusion }of
$A$ in $B$ is a fuzzy relation, denoted by $I(A,B)$. The value of $I(A,B)$ is
defined for each $u\in U$ by
\[
I_{u}(A,B)\doteq\left\{
\begin{tabular}
[c]{ll}%
1 & iff $A_{u}\leq B_{u}$\\
0 & else.
\end{tabular}
\right.
\]
\end{definition}

\begin{example}
\label{exa05}Continuing with the sets of the previous examples, we
have:\ $I(A,C)$ = $[1,1,1,1]$, $I(A,B)$ = $[1,1,1,0]$, $I_{1}(A,B)$ = $1$,
$I_{4}(A,B)$ = $0$. Also, $I(A,\Phi)$ = $[1,1,1,1]$, $I(\Theta,A)$ =
$[1,0,0,0]$, and $I(B,A)$ = $[0,0,0,1]$. Note that $I(A,B)\leq I(A,C)$, but
$I(\Theta,A)||I(B,A)$.
\end{example}

The motivation behind the above definition is as follows. Inclusion of set $A
$ into set $B$ is defined in terms of the elementwise values of $A$ and $B$.
For example, if for every $u\in U$ it is true that $u$ is contained in $B$
more than it is contained in $A$, then we conclude that $A$ is included in $B
$ in the maximum degree, namely \underline{1}. The difference of Definition
\ref{def0301} from alternative definitions of inclusion measure is the
following: traditionally the ``elementwise'' inclusion of two sets is
aggregated into a single real number, i.e. the range of a traditional
inclusion measure is usually totally ordered; whereas, in this work the
elementwise inclusions are collected into $I(A,B)$, a quantity which takes
values in the partially ordered set $\mathbf{B}$ (i.e. a $L$-fuzzy quantity).
Since the values of $I(.,.)$ are partially but not totally ordered, there may
exist sets $X,Y,Z,W$ such that: $X$ is included in $Z$ more than it is
included in $Y$ \ (i.e. $I(X,Y)\leq I(X,Z)$) but the inclusion of $X$ in $Y$
\emph{cannot be compared }to the inclusion of $X$ in $W$ (i.e.
$I(X,Y)||I(X,W)$, viz. the last part of Example \ref{exa05} above).

As mentioned in the introduction, we will occasionally refer to quantities
which take values in a totally ordered set as \emph{``scalar''} and to
quantities which take take values in a partially ordered set as
\emph{``vector''}. The rationale is rather obvious:\ in case $U$ has finite
cardinality ($|U|=N<\infty$), then $I(A,B)$ is indeed a Boolean vector of size
$N$. In case $U$ has infinite (countable or uncountable)\ cardinality,
$I(A,B)$ is a function, which can still be considered as a vector in an
infinite-dimensional vector space. The term ``vector'' is not intended to
invoke connotations of vectorial operations, e.g. addition of inclusion
measures; on the other hand, the partial order $\leq$ $\ $and the $\vee$ and
$\wedge$ operations defined in Section \ref{sec0202} are the ones
``naturally'' introduced for vectors. The term ``scalar'' is used in
contradistinction to ``vector''.

The following proposition describes what we consider the most attractive
property of $I(.,.)$.

\begin{proposition}
\label{pro0302}$I(.,.)$ is a fuzzy order on $\mathbf{F}$, i.e. it is a
reflexive, antisymmetric and $\wedge$-\emph{transitive }fuzzy relation
\end{proposition}

\begin{proof}
Take any $A,B,C\in\mathbf{F}$. We then have the following.

\begin{description}
\item [1.]\emph{Reflexivity}. We have: ($\forall u\in U:$ $A_{u}\leq
A_{u})\Rightarrow$ ($\forall u\in U:I_{u}(A,A)=1)\Rightarrow$ $I(A,A)=$%
\underline{$1$}.

\item[2.] \emph{Antisymmetry}. $I(A,B)=I(B,A)\Rightarrow$ ($\forall u\in U:$
$I_{u}(A,B)=I_{u}(B,A))$. Take any $u\in U$ and consider two cases. (a)\ If
$I_{u}(A,B)=I_{u}(B,A)=1$, then $A_{u}\leq B_{u}$ and $B_{u}\leq A_{u}$ ,
hence $A_{u}=B_{u}$. (b)\ If $I_{u}(A,B)=I_{u}(B,A)=0$, then $A_{u}>B_{u}$ and
$B_{u}>A_{u}$ which leads to a contradiction. Hence we have that:\ \ ($\forall
u\in U:I_{u}(A,B)=I_{u}(B,A))\Rightarrow$ $I(A,B)=I(B,A)$ = \underline{1}
$\Rightarrow$ $A=B$.

\item[3.] $\wedge$-\emph{transitivity}. We take any $u\in U$ and consider two
cases. (a)\ If $I_{u}(A,C)\wedge I_{u}(C,B)=0$, then clearly $I_{u}(A,B)\geq
I_{u}(A,C)\wedge I_{u}(C,B)$. (b)\ If $I_{u}(A,C)\wedge I_{u}(C,B)=1$, then
$A_{u}\leq C_{u}$ and $C_{u}\leq B_{u}$, which implies $A_{u}\leq B_{u}$ and
so $I_{u}(A,B)=1=I_{u}(A,C)\wedge I_{u}(C,B)$. \ Hence we have
that:\ \ ($\forall u\in U:I_{u}(A,B)\geq I_{u}(A,C)\wedge I_{u}(C,B)$)
$\Rightarrow$ $I(A,B)\geq I(A,C)\wedge I(C,B)$ $\Rightarrow$ $I(A,B)\geq
\sup_{C\in\mathbf{F}}I(A,C)\wedge I(C,B)\blacksquare$
\end{description}
\end{proof}

We now proceed to show that $I(.,.)$ has a number of additional properties,
which several authors have axiomatically postulated as appropriate and/or
desirable for measures of inclusion.

\begin{proposition}
\label{pro0303}For all $A,B,C,D\in\mathbf{F}$ we have:
\end{proposition}

\begin{description}
\item [I1]$I(A,B)=\underline{1}\Leftrightarrow A\leq B$; $I(A,B)=$%
\underline{$0$} $\Leftrightarrow A>B$.

\item[I2] $I(A,A^{\prime})=\underline{0}\Leftrightarrow(\forall u\in U$ we
have $A_{u}>1/2).$

\item[I3] $I(A,B)=I(B^{\prime},A^{\prime}).$

\item[I4] $I(A,B)\vee I(B,A)=\underline{1}.$

\item[I5] $B\leq C\Rightarrow I(A,B)\leq I(A,C).$

\item[I6] $B\leq C\Rightarrow I(C,A)\leq I(B,A).$

\item[I7] $I(A,B)\wedge I(C,D)\leq I(A\wedge C,B\wedge D)\wedge I(A\vee
C,B\vee D)\leq$

\item $\qquad\qquad\qquad\qquad I(A\wedge C,B\wedge D)\vee I(A\vee C,B\vee
D)\leq I(A,B)\vee I(C,D).$

\item[I8] $I(A\vee B,C)=I(A,C)\wedge I(B,C).$

\item[I9] $I(A\wedge B,C)=I(A,C)\vee I(B,C).$

\item[I10] $I(A,B\vee C)=I(A,B)\vee I(A,C).$

\item[I11] $I(A,B\wedge C)=I(A,B)\wedge I(A,C).$

\item[I12] $I(A,B)\leq I(A\wedge C,B\wedge C)\wedge I(A\vee C,B\vee C).$
\end{description}

\begin{proof}
Take any $A,B,C,D\in\mathbf{F}$. In proving properties \textbf{I1--I11}, we
will frequently make use of the fact that: for any $u\in U$ the range of
$A_{u}$, $B_{u}$ etc. is [0,1], which is a \emph{totally ordered }set. This
implies, for instance, that either $A_{u}\leq B_{u}$ or $A_{u}>B_{u}$ (which
is generally not true in a partially ordered set); this and similar facts will
be of key importance in several of the arguments presented in the sequel.

\begin{description}
\item [I1]Assume $I(A,B)=\underline{1}$. Then we have: ($\forall u\in U$ :
$I_{u}(A,B)=1)\Leftrightarrow$ $\ (\forall u\in U:A_{u}\leq B_{u}%
)\Leftrightarrow$ $A\leq B$. In a similar manner we can prove $I(A,B)=$%
\underline{$0$} $\Leftrightarrow A>B$.

\item[I2] Assume $I(A,A^{\prime})=$\underline{$0$}. Then for all $u\in U$ we
have $I_{u}(A,A^{\prime})=0\Leftrightarrow$ $A_{u}>A_{u}^{\prime
}\Leftrightarrow$ $A_{u}>1-A_{u}\Leftrightarrow$ $A_{u}>1/2$.

\item[I3] We take any $u\in U$ and consider two cases. (a)\ For all $u\in U$
such that $I_{u}(A,B)=1$ we have $A_{u}\leq B_{u}\Rightarrow B_{u}^{\prime
}\leq A_{u}^{\prime}\Rightarrow I_{u}(B^{\prime},A^{\prime})=1$. (b)\ For all
$u\in U$ such that $I_{u}(A,B)=0$ we have $A_{u}>B_{u}\Rightarrow
B_{u}^{\prime}>A_{u}^{\prime}\Rightarrow I_{u}(B^{\prime},A^{\prime})=0$. So,
we have ($\forall u\in U:$ $I_{u}(A,B)=I_{u}(B^{\prime},A^{\prime})$)
$\Rightarrow$ $I(A,B)=I(B^{\prime},A^{\prime})$.

\item[I4] We take any $u\in U$. Then either $A_{u}\leq B_{u}$ or $A_{u}>B_{u}
$; so either $I_{u}(A,B)=1$ or $I_{u}(B,A)=1$. Hence we have:\ ($\forall u\in
U:$ $I_{u}(A,B)\vee$ $I_{u}(B,A)=1)\Rightarrow$ $I(A,B)\vee$ $I(B,A)=
$\underline{$1$}.

\item[I5] Assume $B\leq C$. Then for all $u\in U$ we have $B_{u}\leq C_{u}$.
Now we take any $u\in U$ and consider two cases. (a) If $A_{u}>B_{u}$, then
$I_{u}(A,B)=0\leq I_{u}(A,C)$. (b) If $A_{u}\leq B_{u}$, then $A_{u}\leq
C_{u}\Rightarrow I_{u}(A,B)=1=I_{u}(A,C)$. Hence we have:\ ($\forall u\in U:$
$I_{u}(A,B)\leq I_{u}(A,C)$) $\Rightarrow$ $I(A,B)\leq I(A,C)$.

\item[I6] Assume $B\leq C$. For all $u\in U$ we have $B_{u}\leq C_{u}$. Now we
take any $u\in U$ and consider two cases.\ \ (a) $C_{u}>A_{u}$, then
$I_{u}(C,A)=0\leq I_{u}(B,A)$; (b) $C_{u}\leq A_{u}$, then $B_{u}\leq
A_{u}\Rightarrow I_{u}(C,A)=1=I_{u}(B,A)$. Hence we have:\ ($\forall u\in U:$
$I_{u}(C,A)\leq I_{u}(B,A)$) $\Rightarrow$ $I(C,A)\leq I(B,A)$.

\item[I7] Take any $u\in U$. If $I_{u}(A,B)=0$ or $I_{u}(C,D)=0$ then
$I_{u}(A,B)\wedge I_{u}(C,D)$ = 0 $\leq$ $I_{u}(A\wedge C,B\wedge D)\wedge
I_{u}(A\vee C,B\vee D)$. If, on the other hand, $I_{u}(A,B)=I_{u}(C,D)=1$ ,
then
\[
\left.
\begin{array}
[c]{c}%
A_{u}\leq B_{u}\\
C_{u}\leq D_{u}%
\end{array}
\right\}  \Rightarrow\left\{
\begin{array}
[c]{c}%
A_{u}\wedge C_{u}\leq B_{u}\wedge D_{u}\\
A_{u}\vee C_{u}\leq B_{u}\vee D_{u}%
\end{array}
\right\}  \Rightarrow\left\{
\begin{array}
[c]{c}%
I_{u}(A\wedge C,B\wedge D)=1\\
I_{u}(A\vee C,B\vee D)=1.
\end{array}
\right.
\]
Hence we have:\ ($\forall u\in U:$ $I_{u}(A,B)\wedge I_{u}(C,D)\leq$
$I_{u}(A\wedge C,B\wedge D)\wedge I_{u}(A\vee C,B\vee D)$) $\Rightarrow$
\begin{equation}
I(A,B)\wedge I(C,D)\leq I(A\wedge C,B\wedge D)\wedge I(A\vee C,B\vee D).
\label{eq01}%
\end{equation}
Clearly we also have
\begin{equation}
I(A\wedge C,B\wedge D)\wedge I(A\vee C,B\vee D)\leq I(A\wedge C,B\wedge D)\vee
I(A\vee C,B\vee D). \label{eq02}%
\end{equation}
Finally, take any $u\in U$. If $I_{u}(A,B)\vee$ $I_{u}(C,D)$ =1, then
obviously $I_{u}(A\wedge C,B\wedge D)\vee I_{u}(A\vee C,B\vee D)$ $\leq$
$I_{u}(A,B)\vee$ $I_{u}(C,D)$. If, on the other hand, $I_{u}(A,B)\vee$
$I_{u}(C,D)$ =0, then $I_{u}(A,B)=I_{u}(C,D)=0$ , and so
\[
\left.
\begin{array}
[c]{c}%
A_{u}>B_{u}\\
C_{u}>D_{u}%
\end{array}
\right\}  \Rightarrow\left\{
\begin{array}
[c]{c}%
A_{u}\wedge C_{u}>B_{u}\wedge D_{u}\\
A_{u}\vee C_{u}>B_{u}\vee D_{u}%
\end{array}
\right\}  \Rightarrow\left\{
\begin{array}
[c]{c}%
I_{u}(A\wedge C,B\wedge D)=0\\
I_{u}(A\vee C,B\vee D)=0.
\end{array}
\right.
\]
Hence we have:\ ($\forall u\in U:$ $I_{u}(A\wedge C,B\wedge D)\vee I_{u}(A\vee
C,B\vee D)$ $\leq$ $I_{u}(A,B)\vee I_{u}(C,D)$) $\Rightarrow$
\begin{equation}
I(A\wedge C,B\wedge D)\vee I(A\vee C,B\vee D)\leq I(A,B)\vee I(C,D).
\label{eq03}%
\end{equation}
Now \textbf{I7} is immediately obtained from (\ref{eq01}-\ref{eq03}).

\item[I8] In (\ref{eq01}) substitute $B$ with $C$, $C$ with $B$ and $D$ with
$C$, to obtain
\begin{equation}
I(A,C)\wedge I(B,C)\leq I(A\wedge B,C)\wedge I(A\vee B,C)\leq I(A\vee B,C).
\label{eq11}%
\end{equation}
On the other hand, from \textbf{I6} we have
\begin{equation}
\left.
\begin{tabular}
[c]{l}%
$A\leq A\vee B\Rightarrow I(A\vee B,C)\leq I(A,C)$\\
$B\leq A\vee B\Rightarrow I(A\vee B,C)\leq I(B,C)$%
\end{tabular}
\right\}  \Rightarrow I(A\vee B,C)\leq I(A,C)\wedge I(B,C). \label{eq12}%
\end{equation}
Now (\ref{eq11},\ref{eq12}) yield the desired result.

\item[I9] In (\ref{eq03}) substitute $B$ with $C$, $C$ with $B$ and $D$ with
$C$, to obtain
\begin{equation}
I(A\wedge B,C)\leq I(A\wedge B,C)\vee I(A\vee B,C)\leq I(A,C)\vee I(B,C).
\label{eq21}%
\end{equation}
On the other hand, from \textbf{I5} we have
\begin{equation}
\left.
\begin{tabular}
[c]{l}%
$A\wedge B\leq A\Rightarrow I(A,C)\leq I(A\wedge B,C)$\\
$A\wedge B\leq B\Rightarrow I(B,C)\leq I(A\wedge B,C)$%
\end{tabular}
\right\}  \Rightarrow I(A,C)\vee I(B,C)\leq I(A\wedge B,C). \label{eq22}%
\end{equation}
Now (\ref{eq21},\ref{eq22}) yield the desired result.

\item[I10] In (\ref{eq03}) substitute $C$ with $A$ and $D$ with $C$, to
obtain
\begin{equation}
I(A,B\vee C)\leq I(A,B\wedge C)\vee I(A,B\vee C)\leq I(A,B)\vee I(A,C).
\label{eq31}%
\end{equation}
On the other hand, from \textbf{I5} we have
\begin{equation}
\left.
\begin{tabular}
[c]{l}%
$B\leq B\vee C\Rightarrow I(A,B)\leq I(A,B\vee C)$\\
$C\leq B\vee C\Rightarrow I(A,C)\leq I(A,B\vee C)$%
\end{tabular}
\right\}  \Rightarrow I(A,B)\vee I(A,C)\leq I(A,B\vee C). \label{eq32}%
\end{equation}
Now (\ref{eq31},\ref{eq32}) yield the desired result.

\item[I11] In (\ref{eq01}) substitute $C$ with $A$ and $D$ with $C$, to
obtain
\begin{equation}
I(A,B)\wedge I(A,C)\leq I(A,B\wedge C)\wedge I(A,B\vee C)\leq I(A,B\wedge C).
\label{eq41}%
\end{equation}
On the other hand, from \textbf{I6} we have
\begin{equation}
\left.
\begin{tabular}
[c]{l}%
$B\wedge C\leq B\Rightarrow I(A,B\wedge C)\leq I(A,B)$\\
$B\wedge C\leq C\Rightarrow I(A,B\wedge C)\leq I(A,C)$%
\end{tabular}
\right\}  \Rightarrow I(A,B\wedge C)\leq I(A,B)\wedge I(A,C). \label{eq42}%
\end{equation}
Now(\ref{eq41},\ref{eq42}) yield the desired result.

\item[I12] In (\ref{eq01}) substitute $D$ with $C$ to get
\begin{equation}
I(A,B)\wedge I(C,C)\leq I(A\wedge C,B\wedge C)\wedge I(A\vee C,B\vee C).
\label{eq43}%
\end{equation}
But $I(C,C)=$\underline{$1$} and so $I(A,B)\wedge I(C,C)$ = $I(A,B).$ Now
(\ref{eq43}) yields \textbf{I12} immediately.$\blacksquare$
\end{description}
\end{proof}

\noindent\textbf{Remark}. Properties \textbf{I5, I6} express the monotonicity
of $I(.,.)$ in the second and first argument, respectively. Properties
\textbf{I8--I11} are related to the combination of this monotonicity and
$\wedge$--transitivity; it is worth remarking that for inclusion measures
which do not enjoy $\wedge$--transitivity, the corresponding properties are
weaker, with $\leq$ or $\geq$ in place of = (compare with the properties
listed in Section \ref{sec0302}).

\begin{proposition}
\label{pro0304}For all $A,B,C,D\in\mathbf{F}$ and all $\Theta\in\mathbf{B}$ we
have:$\;$%
\[
I(A,B)\geq\Theta\Rightarrow I(A\wedge C,B\wedge C)\wedge I(A\vee C,B\vee
C)\geq\Theta.
\]
\end{proposition}

\begin{proof}
This follows immediately from \textbf{I12}.$\ \blacksquare$
\end{proof}

\subsection{Comparison to Previously Proposed Inclusion Measures}

\label{sec0302}

In this section we briefly compare our inclusion measure to the ones
introduced by other authors. We will list some of the axioms/properties
proposed by several authors as appropriate for a reasonable inclusion measure
and will compare these to the properties of $I(.,.)$ introduced in this paper.
We consider three axiomatizations of inclusion measure properties:\ Sinha and
Dougherty's (\cite{Sinh01}, also reproduced in \cite{Bur01}), Fan's
\cite{JFan01} and Young's \cite{Virg01}. We present these axioms in Table 1.
However, before proceeding some remarks are in \ order.

First of all, let us note that we have adapted the notation of the authors
mentioned above so as to parallel our notation. In particular, we have used
the symbol $i(.,.)$ to denote a scalar inclusion measure (in contradistinction
to our vector $I(.,.)$); also, since the axioms/ properties appearing in Table
1 refer to inclusion measures with a totally ordered range, the maximum
element is 1 rather than \underline{1} and the minimum element is 0 rather
than \underline{0}.

Second, we must remark that some of the authors previously mentioned present
several alternative lists of axioms/ properties; hence Table 1 is meant to
give a representative (but not exclusive) view of what are considered
``reasonable'' properties for an inclusion measure.

Finally, let us note that the term ``axiom'' (even though used by some
authors) is not entirely appropriate, because some of the properties listed in
Table 1 are not independent of each other. For instance, \textbf{i4a} implies
\textbf{i8} and \textbf{i5a} implies \textbf{i7}. Also, it can be seen that
\textbf{i4a} is a stronger form of \textbf{i4b} and \textbf{i5a} is a stronger
form of \textbf{i5b.} \ Similarly, we have altered the order of presentation
of the axioms/ properties so as to agree with the one we have used.

The first column in Table 1 lists an id. number and the the final column lists
the corresponding property of our inclusion measure, as appearing in
Proposition \ref{pro0303}. The third column list the actual axiom / property
and the remaining columns indicate the papers which use the particular property.

\begin{center}%
\begin{tabular}
[c]{|l|l|l|l|l|l|}\hline
No. & Property & \cite{Bur01,Sinh01} & \cite{JFan01} & \cite{Virg01} &
Corresponding\\
&  &  &  &  & in this paper\\\hline
\textbf{i1} & $A\leq B\Leftrightarrow i(A,B)=1$ & $\times$ & $\times$ &
$\times$ & \textbf{I1}\\\hline
\textbf{i2a} & $i(A,A^{\prime})=0\Leftrightarrow A$ is crisp set &  &  &
$\times$ & \textbf{I2}\\
\textbf{i2b} & ($\forall u\in U:\frac{1}{2}\leq A_{u})\Rightarrow
(i(A,A^{\prime})=0\Leftrightarrow A=U)$ &  & $\times$ &  & \textbf{I2}\\\hline
\textbf{i3} & $i(A,B)=i(B^{\prime},A^{\prime})$ & $\times$ & $\times$ &  &
\textbf{I3}\\\hline
\textbf{i4a} & $B\leq C\Rightarrow i(A,B)\leq i(A,C)$ & $\times$ &  & $\times$%
& \textbf{I5}\\
\textbf{i4b} & $B\leq C\leq A\Rightarrow i(A,B)\leq i(A,C)$ &  & $\times$ &  &
\textbf{I5}\\\hline
\textbf{i5a} & $B\leq C\Rightarrow i(C,A)\leq i(B,A)$ & $\times$ &  & $\times$%
& \textbf{I6}\\
\textbf{i5b} & $A\leq B\leq C\Rightarrow i(C,A)\leq i(B,A)$ &  & $\times$ &  &
\textbf{I6}\\\hline
\textbf{i6} & $i(A\vee B,C)=i(A,C)\wedge i(B,C)$ & $\times$ &  &  &
\textbf{I8}\\\hline
\textbf{i7} & $i(A\wedge B,C)\geq i(A,C)\vee i(B,C)$ & $\times$ &  &  &
\textbf{I9}\\\hline
\textbf{i8} & $i(A,B\vee C)\geq i(A,B)\vee i(A,C)$ & $\times$ &  &  &
\textbf{I10}\\\hline
\textbf{i9} & $i(A,B\wedge C)=i(A,B)\wedge i(A,C)$ & $\times$ &  &  &
\textbf{I11}\\\hline
\textbf{i10} & $i(A,B)=0\Leftrightarrow\exists u\in U:A_{u}=1,B_{u}=0$ &
$\times$ &  &  & \textbf{I1}\\\hline
\end{tabular}

\textbf{Table 1:} Inclusion Measure Axioms
\end{center}

It can be seen from the above table that for every property appearing in Table
1, our $I(.,.)$ enjoys an analogous (and sometimes stronger) property. While
Table 1 is not an exhaustive list of all properties proposed as reasonable for
an inclusion measure, it offers a pretty complete coverage of such properties;
the ones omitted are not, in our judgement, crucial\footnote{For instance, we
have omitted Sinha and Dougherty's property about invariance of inclusion
measure with respect to the shift operation; but a similar property can be
easily proved with respect to relabelings of set elements.}. In this
connection, Sinha and Dougherty's remarks (pp.19-20 of \cite{Sinh01})
regarding the choice of appropriate ``axioms'' are particularly relevant.

In conclusion, our $I(.,.)$\ appears to be a rather reasonable inclusion
measure in the sense that it has a larg enumber of reasonable properties.

In addition $I(.,.)$ enjoys $\wedge$--transitivity. This is usually \emph{not}
the case with the scalar inclusion measures appearing in the literature. This
can be seen by considering the following list of specific examples of such
inclusion measures. In the following list we use the popular notation
$|A|=\sum_{u\in U}A_{u}$.

\begin{enumerate}
\item $i(A,B)=|A\wedge B|/|A|$.

\item $i(A,B)=|B|/|A\vee B|.$

\item $i(A,B)=|A^{\prime}\wedge B^{\prime}|/|B^{\prime}|.$

\item $i(A,B)=|A^{\prime}|/|A^{\prime}\vee B^{\prime}|.$

\item $i(A,B)=|A^{\prime}\vee B|/|A^{\prime}\vee A\vee B\vee B^{\prime}|.$

\item $i(A,B)=|A^{\prime}\wedge A\wedge B\wedge B^{\prime}|/|A\wedge
B^{\prime}|.$

\item $i(A,B)=(|A^{\prime}|\vee|B|)/|A^{\prime}\vee A\vee B\vee B^{\prime}|. $

\item $i(A,B)=|A^{\prime}\wedge A\wedge B\wedge B^{\prime}|/(|A|\wedge
|B^{\prime}|).$

\item $i(A,B)=\frac{\sum_{u\in U}1\wedge(1-A_{u}+B_{u})}{|U|}.$

\item $i(A,B)=$ $\frac{\sum_{u\in U}(1-A_{u})\vee B_{u}}{|U|}.$

\item $i(A,B)=\frac{\sum_{u\in U}(1-A_{u}+A_{u}B_{u})}{|U|}.$

\item $i(A,B)=\sup\{\alpha:\forall u\in U$ we have $A_{u}\wedge\alpha\leq
B_{u}\wedge\alpha\}$.
\end{enumerate}

Inclusion measure no.1 above is Kosko's inclusion measure \cite{Kosko1}; no.
12 is introduced in \cite{Kundu1} and does not appear very widely in the
literature; for references to the remaining inclusion measures see
\cite{JFan01} and \cite{Virg01}.

The only inclusion measure in the above list which is $\wedge$-transitive is
no.12, Kundu's inclusion measure. In addition, Willmott introduces some
transitive fuzzy inclusion measures in \cite{Willmott1}, which however involve
a rather drastically modified definition of transitivity. In conclusion, it
appears that $\wedge$--transitivity is a requirement which cannot be easily
satisfied by an inclusion measure which takes values in a totally ordered
range (``scalar'' inclusion measure). On the other hand it is satisfied (in
addition with a large number of other properties)\ by our L-fuzzy inclusion measure.

\section{Similarity and Distance Defined in Terms of L-Fuzzy Inclusion}

\label{sec04}

We now proceed to define L-fuzzy similarity and L-fuzzy distance in terms of
L-fuzzy inclusion.

\subsection{L-Fuzzy Similarity}

\label{sec0401}

\begin{definition}
\label{def0401}For all $A,B\in\mathbf{F}$, the \emph{measure of similarity
}between $A$ and $B$ \ is denoted by $S(A,B)$ and is defined by
\[
S(A,B)\doteq I(A,B)\wedge I(B,A).
\]
\end{definition}

The domain of $S(.,.)$ is the partially ordered set $\mathbf{B}$, and this
results to possibly incomparable similarity measures.

\begin{example}
\label{exa06}Continuing with the sets of the previous examples, we
have:\ $S(\Phi,\Theta)$ = $I(\Phi,\Theta)\wedge I(\Theta,\Phi)$ =
$[0,1,1,1]\wedge\lbrack1,1,0,1]$ = $[0,1,0,1]$; and $S(\Phi,A)$ =
$I(\Phi,A)\wedge I(A,\Phi)$ = $[0,0,1,0]\wedge\lbrack1,1,1,1]$ = $[0,0,1,0]$.
\end{example}

The rationale of the above definition is as follows. If for two sets $A$ and
$B$ we have that $A$ is included in $B$ to the maximum degree and vice versa,
then the sets are identical and have maximum similarity. In fact, the basic
property is given by the following proposition.

\begin{proposition}
For all $A,B\in\mathbf{F}$ and for all $u\in U$ we have:$\;S_{u}%
(A,B)=1\Leftrightarrow A_{u}=B_{u}$.
\end{proposition}

\begin{proof}
Take any $A,B\in\mathbf{F}$ and any $u\in U$. $S_{u}(A,B)=1$ $\Leftrightarrow
(I_{u}(A,B)=1\ $ and $I_{u}(B,A)=1)$ $\ \Leftrightarrow(A_{u}\leq B_{u}\ $ and
$B_{u}\leq A_{u})\Leftrightarrow$ $A_{u}=B_{u}$.$\blacksquare$
\end{proof}

Further properties are presented in Propositions \ref{pro0402}, \ref{pro0403},
\ref{pro0404}.

\begin{proposition}
\label{pro0402}$S(.,.)$ is a fuzzy similarity relation on $\mathbf{F}$, i.e.
it is reflexive, symmetric and $\wedge$-\emph{transitive}.
\end{proposition}

\begin{proof}
Take any $A,B,C\in\mathbf{F}$. Then we have the following.

\begin{enumerate}
\item \emph{Reflexivity}. From Proposition \ref{pro0302} we have that
$I(A,A)=$\underline{$1$}. Since $S(A,A)=I(A,A)\wedge I(A,A)$ it follows
$S(A,A)=$\underline{$1$}.

\item \emph{Symmetry}. $S(A,B)=I(A,B)\wedge I(B,A)=I(B,A)\wedge I(A,B)=S(B,A)
$.

\item $\wedge$\emph{-transitivity}. Set $\Theta\doteq I(A,B)$, $\Psi\doteq
I(B,A)$, $\Phi\doteq I(A,C)$, $\Gamma\doteq I(C,A)$, $\Omega\doteq I(C,B)$,
$\Delta\doteq I(B,C)$. From Proposition \ref{pro0302} we have $\Theta\geq
\Phi\wedge\Omega$, $\Psi\geq\Delta\wedge\Gamma$. Hence $\Theta\wedge\Psi
\geq(\Phi\wedge\Omega$) $\wedge$ ($\Gamma\wedge\Delta$) $\Rightarrow
\Theta\wedge\Psi\geq$ $(\Phi\wedge\Gamma$) $\wedge$ ($\Omega\wedge\Delta$)
$\Rightarrow$ $S(A,B)\geq S(A,C)\wedge S(C,B)\Rightarrow$ $S(A,B)\geq
\sup_{C\in\mathbf{F}}(A,C)\wedge S(C,B)$.$\blacksquare$
\end{enumerate}
\end{proof}

\begin{proposition}
\label{pro0403}For all $A,B,C,D\in\mathbf{F}$ we have:

\begin{description}
\item [S1]$S(A,B)=\underline{1}\Leftrightarrow A=B$; $S(A,B)=\underline
{0}\Rightarrow A\neq B$.

\item[S2a] $S(A,A^{\prime})=\underline{0}\Leftrightarrow(\forall u\in U$ we
have $A_{u}\neq1/2)$.

\item[S2b] $S(A,A^{\prime})=\underline{1}\Leftrightarrow(\forall u\in U$ we
have $A_{u}=1/2)$.

\item[S3] $S(A,B)=S(A^{\prime},B^{\prime}).$

\item[S4] $A\leq B\leq C\Rightarrow\left\{
\begin{tabular}
[c]{l}%
$S(A,B)\geq S(A,C)$\\
$S(B,C)\geq S(A,C)$\\
$S(A,C)=S(A,B)\wedge S(B,C).$%
\end{tabular}
\right.  $

\item[S5] $S(A,B)\wedge S(C,D)\leq\left\{
\begin{tabular}
[c]{l}%
$S(A\vee C,B\vee D)$\\
$S(A\wedge C,B\wedge D).$%
\end{tabular}
\right.  $

\item[S6] $S(A,B)\wedge S(A,C)\leq\left\{
\begin{tabular}
[c]{l}%
$S(A,B\vee C)$\\
$S(A,B\wedge C).$%
\end{tabular}
\right.  $

\item[S7] $S(A,C)\vee S(B,C)\geq\left\{
\begin{tabular}
[c]{l}%
$S(A\vee B,C)$\\
$S(A\wedge B,C).$%
\end{tabular}
\right.  $

\item[S8] $S(A,B)\leq\left\{
\begin{tabular}
[c]{l}%
$S(A\vee C,B\vee C)$\\
$S(A\wedge C,B\wedge C).$%
\end{tabular}
\right.  $

\item[S9a] $S(A,A\vee B)=S(B,A\wedge B).$

\item[S9b] $S(A,A\wedge B)=S(B,A\vee B).$

\item[S10] All of the following quantities are equal:
\[%
\begin{array}
[c]{llll}%
(i) & S(A,B)\qquad & (iv) & S(A\wedge B,A\vee B)\qquad\\
(ii) & S(A\wedge B,A)\wedge S(A,A\vee B)\qquad & (v) & S(A,A\vee B)\wedge
S(B,A\vee B)\qquad\\
(iii) & S(A\wedge B,B)\wedge S(B,A\vee B) & (vi) & S(A,A\wedge B)\wedge
S(B,A\wedge B)
\end{array}
.
\]
\end{description}
\end{proposition}

\begin{proof}
Take any $A,B,C,D\in\mathbf{F}$. Then we have the following.

\begin{description}
\item [S1]$S(A,B)=$\underline{$1$}$\Leftrightarrow$ $\ I(A,B)=I(B,A)=$%
\underline{$1$}$\Leftrightarrow$ $(I(A,B)=\underline{1}$ and
$I(B,A)=\underline{1})\ \Leftrightarrow\ (A\leq B$ and $B\leq
A)\Leftrightarrow$ $\ A=B.$ Similarly we can prove $S(A,B)=\underline
{0}\Rightarrow A\neq B$.

\item[S2a] $S(A,A^{\prime})=$\underline{$0$}$\Rightarrow$ $\ (\forall$ $u\in
U$ : $I_{u}(A,A^{\prime})\wedge I_{u}(A^{\prime},A)=0)\Rightarrow$
\ $\ (\forall$ $u\in U$ : $A_{u}>1-A_{u}$ or $1-A_{u}>A_{u})\Rightarrow$
($\forall$ $u\in U$ : $A_{u}\neq1/2$ . Conversely, assume that for all $u\in
U$ we have $A_{u}\neq1/2$. \ It is easy to see that if, for example,
$A_{u}<1/2$ then $I_{u}(A^{\prime},A)=0$; similarly if $A_{u}>1/2$ then
$I_{u}(A,A^{\prime})=0$. Hence we have: ($\forall u\in U:$ $I_{u}(A^{\prime
},A)\wedge I(A,A^{^{\prime}})=0)\Rightarrow$ ($\forall u\in U:S_{u}%
(A,A^{^{\prime}})=0$) $\Rightarrow$ $S(A,A^{\prime})=$\underline{$0 $}$.$

\item[S2b] $S(A,A^{\prime})=$\underline{$1$}$\Leftrightarrow$ $I(A,A^{\prime
})\wedge I(A^{\prime},A)=$\underline{$1$}$\Leftrightarrow$ ($\forall$ $u\in U$
: $I_{u}(A,A^{\prime})=I_{u}(A^{\prime},A)=1)\Leftrightarrow$ ($\forall$ $u\in
U$ : $A_{u}=A_{u}^{\prime}=1/2).$ \ 

\item[S3] $S(A,B)=I(A,B)\wedge I(B,A)$ = $\ I(B^{\prime},A^{\prime})\wedge
I(A^{\prime},B^{\prime})$ = $S(B^{\prime},A^{\prime})$ = $S(A^{\prime
},B^{\prime})$.

\item[S4] Take any $u\in U$ and consider two cases. (a)\ $S_{u}(A,C)=0\leq
S_{u}(A,B)$. (b)\ \ $S_{u}(A,C)=1$. Then $A_{u}=C_{u}$ and this, in
conjunction to $A_{u}\leq B_{u}\leq C_{u}$, implies $A_{u}=B_{u}\Rightarrow$
$S_{u}(A,B)=1$ = $S_{u}(A,C).$ Hence we have: ($\forall u\in U:$
$S_{u}(A,B)\geq S_{u}(A,C)$) $\Rightarrow$ $S(A,B)\geq S(A,C)$. It can be
proved similarly that $S(B,C)\geq S(A,C)$. Finally, from $S(A,B)\geq S(A,C)$
and $S(B,C)\geq S(A,C)$ follows that $S(A,B)\wedge$ $S(B,C)\geq S(A,C)$; but
from transitivity we also have $S(A,B)\wedge$ $S(B,C)\leq S(A,C)$ and so
$S(A,B)\wedge$ $S(B,C)=S(A,C)$.

\item[S5] From Proposition \ref{pro0303}, \textbf{I7} we have
\[
\left.
\begin{tabular}
[c]{l}%
$I(A,B)\wedge I(C,D)\leq I(A\wedge C,B\wedge D)$\\
$I(B,A)\wedge I(D,C)\leq I(B\wedge D,A\wedge C)$%
\end{tabular}
\right\}  \Rightarrow
\]
\[
I(A,B)\wedge I(C,D)\wedge I(B,A)\wedge I(D,C)\leq I(A\wedge C,B\wedge D)\wedge
I(B\wedge D,A\wedge C)\Rightarrow
\]
\[
S(A,B)\wedge S(C,D)\leq S(A\wedge C,B\wedge D).
\]
Similarly we can prove $S(A,B)\wedge S(C,D)\leq S(A\vee C,B\vee D)$.

\item[S6] This is obtained from \textbf{S5} by substituting $C$ with $A$ and
$D$ with $C.$

\item[S7] From Proposition \ref{pro0303}, \textbf{I8} and \textbf{I10} imply
\[
\left.
\begin{tabular}
[c]{l}%
$I(A\vee B,C)=I(A,C)\wedge I(B,C)$\\
$I(C,A\vee B)=I(C,A)\vee I(C,B)$%
\end{tabular}
\right\}  \Rightarrow
\]
\[
S(A\vee B,C)=I(A\vee B,C)\wedge I(C,A\vee B)=(I(A,C)\wedge I(B,C))\wedge
(I(C,A)\vee I(C,B))=
\]
\[
(I(A,C)\wedge I(B,C)\wedge I(C,A))\vee(I(A,C)\wedge I(B,C)\wedge I(C,B))\leq
\]
\[
(I(A,C)\wedge(I(C,A)\vee(I(B,C)\wedge I(C,B))=S(A,C)\vee S(B,C).
\]
It can be proved similarly that $S(A,C)\vee S(B,C)\geq S(A\wedge B,C)$.

\item[S8] This follows from \textbf{S5}, using $C=D$ and noting that
$S(A,B)\wedge S(C,C)=S(A,B)\wedge\underline{1}=S(A,B)$.

\item[S9] Choose any $u\in U$ and consider two cases. (a)\ If $S_{u}(A,A\vee
B)=1$, then $A_{u}=A_{u}\vee B_{u}\Rightarrow$ \ $A_{u}\geq B_{u}\Rightarrow$
$S_{u}(B,A\wedge B)=1$. (b) If $S_{u}(A,A\vee B)=0$, then $A_{u}\neq A_{u}\vee
B_{u}\Rightarrow$ $A_{u}<B_{u}\Rightarrow$ $A_{u}\wedge B_{u}$ = $A_{u}%
<B_{u}\Rightarrow$ \ $S_{u}(B,A\wedge B)=0$. Hence, for all $u\in U$ we have
$S_{u}(A,A\vee B)$ = $S_{u}(B,A\wedge B)$, i.e. $S(A,A\vee B)$ = $S(B,A\wedge
B)$. By interchanging the role of $A$ and $B$, we obtain $S(B,A\vee B)$ =
$S(A,A\wedge B)$.

\item[S10] We will prove this by showing that (i) is equal to each of (ii)\ -- (vi).

\begin{description}
\item [(i)=(ii)]From symmetry and transitivity we have
\[
S(A,B)=S(B,A)\geq S(B,A\vee B)\wedge S(A\vee B,A)\Rightarrow
\]
(by use of \textbf{S9b} and symmetry)
\begin{equation}
S(A,B)\geq S(A,A\wedge B)\wedge S(A,A\vee B). \label{eq55}%
\end{equation}
Also, if in \textbf{S6} we substitute $B$ with $A$ and $C$ with $B,$ we obtain
(using $S(A,A)\wedge S(A,B)$ = \underline{$1$}$\wedge S(A,B)$ =$S(A,B) $)
\[
S(A,A)\wedge S(A,B)\leq\left\{
\begin{tabular}
[c]{l}%
$S(A,A\vee B)$\\
$S(A,A\wedge B)$%
\end{tabular}
\right\}  \Rightarrow
\]
\begin{equation}
S(A,B)\leq S(A,A\wedge B)\wedge S(A,A\vee B). \label{eq56}%
\end{equation}
From (\ref{eq55},\ref{eq56}) we see that $S(A,B)=S(A,A\wedge B)\wedge
S(A,A\vee B)$ = $S(A\wedge B,A)\wedge S(A,A\vee B).$

\item[(i)=(iii)] This is proved exactly as above, interchanging the role of
$A$ and $B$.

\item[(i)=(iv)] We have $A\wedge B\leq A\leq A\vee B$. Using \textbf{S4} and
that (i)=(ii)\textbf{, }we immediately get $S(A\wedge B,A\vee B)$ = $S(A\wedge
B,A)\wedge S(A,A\vee B)$ = $S(A,B)$.

\item[(i)=(v)] We have (using \textbf{S9b} and symmetry)
\begin{equation}
S(A,A\vee B)\wedge S(B,A\vee B)=S(A,A\vee B)\wedge S(A,A\wedge B)=S(A,A\vee
B)\wedge S(A\wedge B,A) \label{eq57}%
\end{equation}
but we have already proved that $S(A,B)$ = $S(A\wedge B,A)\wedge S(A,A\vee
B)\ $ and so the proof is complete.

\item[(i)=(vi)] This proved similarly to the previous step.
\end{description}

\item  From the last few steps we see that (i) = (ii) = ... = (vi). This
completes the proof of \textbf{S10} and of the proposition.$\blacksquare$
\end{description}
\end{proof}

\noindent\textbf{Remark}. Property \textbf{S4 }is related to the concepts of
\emph{betweenness} and \emph{\ convexity}; property \textbf{S10} is related to
modularity. However, since these concepts are usually related to distance,
rather than similarity, we will present the corresponding remarks in Section
\ref{sec0402}.

\begin{proposition}
\label{pro0404}For all $A,B,C\in\mathbf{F}$ and all $\Theta\in\mathbf{B}$ we have:

\begin{enumerate}
\item $S(A,B)\geq\Theta\Rightarrow S(A\vee C,B\vee C)\geq\Theta$.

\item $S(A,B)\geq\Theta\Rightarrow S(A\wedge C,B\wedge C)\geq\Theta$.

\item $S(A\wedge C,B\wedge C)=S(A\vee C,B\vee C)=\Theta\Rightarrow
S(A,B)=\Theta.$
\end{enumerate}
\end{proposition}

\begin{proof}
Choose any $A,B,C\in\mathbf{F}$ and any $\Theta\in\mathbf{B}$. Then we have
the following.

\begin{enumerate}
\item  This follows from \textbf{S8}.

\item  This follows from \textbf{S8}.

\item  Take any $A,B,C\in\mathbf{F}$ and any $\Theta\in\mathbf{B}$ such that
$S(A\wedge C,B\wedge C)=S(A\vee C,B\vee C)=\Theta$. Choose any $u\in U$. We
consider two cases. (a) If $\Theta_{u}=1$, then $A_{u}\wedge C_{u}$ =
$\ B_{u}\wedge C_{u}$ and $A_{u}\vee C_{u}$ = $\ B_{u}\vee C_{u}$ . Then, by
distributivity we have $A_{u}=B_{u}\Rightarrow$ $S_{u}(A,B)$ =1 = $\Theta_{u}
$. (b) If \ $\Theta_{u}=0$, then $A_{u}\wedge C_{u}$ $\neq$ $\ B_{u}\wedge
C_{u}$ and $A_{u}\vee C_{u}$ $\neq$ $\ B_{u}\vee C_{u}$ and so, obviously,
$A_{u}\neq B_{u}\Rightarrow S_{u}(A,B)=0=\Theta_{u}$. Hence, for all $u\in U$
we have $S_{u}(A,B)=\Theta_{u}$.$\blacksquare$
\end{enumerate}
\end{proof}

\noindent\textbf{Remark}. Proposition \ref{pro0404} is related to the concept
of $\epsilon$-similarity introduced by Pappis
\cite{Pappis01,Pappis02,Pappis03}.

\subsection{L-Fuzzy Distance}

\label{sec0402}

\begin{definition}
\label{def0501}For $A,B\in\mathbf{F}$, the distance between $A$ and $B$ \ is
denoted by $D(A,B)$ and is defined by
\[
D(A,B)\doteq S^{\prime}(A,B).
\]
\end{definition}

\begin{example}
Continuing with the sets of the previous examples, we have:\ $D(\Phi,\Theta)$
= $S^{\prime}(\Phi,\Theta)$ = $[0,1,0,1]^{\prime}$ = $[1,0,1,0]$; and
$D(\Phi,A)$ = $S^{\prime}(\Phi,A)$ = $[0,0,1,0]^{\prime}$ = $[1,1,0,1]$.
\end{example}

\begin{proposition}
For all $A,B\in\mathbf{F}$ and for all $u\in U$ we have:$\;D_{u}%
(A,B)=0\Leftrightarrow A_{u}=B_{u}$.
\end{proposition}

\begin{proof}
Take any $A,B\in\mathbf{F}$ and any $u\in U$. $D_{u}(A,B)=0\Leftrightarrow
S_{u}(A,B)=1$ $\Leftrightarrow A_{u}=B_{u}$.$\blacksquare$
\end{proof}

As can be seen by the definition and by the above proposition, distance is
defined as the complement of similarity. Given that complementation is
order-inverting, it follows that if sets $A$ and $B$ have large similarity,
then they will have small distance. Since distance is usually perceived as a
totally ordered, nonnegative quantity, the above definition may appear rather
unusual. In fact however, $D(.,.)$ has the basic characteristics of a distance
function, as outlined in the next proposition.

\begin{proposition}
\label{pro0502}For all $A,B,C\in\mathbf{F}$ we have:

\begin{enumerate}
\item $D(A,B)=\underline{0}$ $\Leftrightarrow$ $A=B$.

\item $D(A,B)=D(B,A)$.

\item $D(A,B)\leq D(A,C)\vee D(C,B)$.
\end{enumerate}
\end{proposition}

\begin{proof}
This follows from the definition $D(A,B)=S^{\prime}(A,B)$ and from Proposition
\ref{pro0402}.$\blacksquare$
\end{proof}

\noindent\textbf{Remark}. $D(A,B)\leq D(A,C)\vee D(C,B)$ is the partial order
analog of the triangle inequality (in addition it is the \emph{ultrametric}
triangle inequality). The idea of Boolean valued distances is not new; it
appears already in the 1950's (for instance see Blumenthal's book
\cite{Blum99}). This idea has been applied to Boolean lattices (which are
called ``\emph{auto}metrized'' spaces since the domain of the metric distance
function is the same space on which the metric is imposed). Formally, in
\cite{Blum99} we have $D:\mathbf{B}\times\mathbf{B}\rightarrow\mathbf{B}$. In
our case we have $D:\mathbf{F}\times\mathbf{F}\rightarrow\mathbf{B\subseteq
F}$; in this sense ($\mathbf{F},D$) is an autometrized space. The following
proposition is also of interest.

\begin{proposition}
\label{pro0503}For all $\Theta,\Phi\in\mathbf{B}$ we have $D(\Theta
,\Phi)=(\Theta^{\prime}\wedge\Phi)\vee(\Theta\wedge\Phi^{\prime})$.
\end{proposition}

\begin{proof}
Take any $\Theta,\Phi\in\mathbf{B}$ and any $u\in U$. By considering the four
cases (a)\ $\Theta_{u}$= 0, $\Phi_{u}$= 0, (b)\ $\Theta_{u}$= 0, $\Phi_{u}$=
1, (c)\ $\Theta_{u}$= 1, $\Phi_{u}$= 0, (d)\ $\Theta_{u}$= 1, $\Phi_{u}$= 1,
we see that in every case (i.e. for all $u\in U$)\ we have $D_{u}(\Theta
,\Phi)$ = $(\Theta_{u}^{\prime}\wedge\Phi_{u})\vee(\Theta_{u}\wedge\Phi
_{u}^{\prime})$ and the proof is complete.$\blacksquare$
\end{proof}

Let us present some further properties of $D(.,.)$.

\begin{proposition}
\label{pro0504}For all $A,B,C,E\in\mathbf{F}$ we have:

\begin{description}
\item [D1]$D(A,B)=\underline{0}\Leftrightarrow A=B$.

\item[D2a] $D(A,A^{\prime})=\underline{1}\Leftrightarrow(\forall u\in U$ we
have $A_{u}\neq1/2)$.

\item[D2b] $D(A,A^{\prime})=\underline{0}\Leftrightarrow(\forall u\in U$ we
have $A_{u}=1/2)$.

\item[D3] $D(A,B)=D(A^{\prime},B^{\prime}).$

\item[D4] $A\leq B\leq C\Rightarrow\left\{
\begin{tabular}
[c]{l}%
$D(A,B)\leq D(A,C)$\\
$D(B,C)\leq D(A,C)$\\
$D(A,C)=D(A,B)\vee D(B,C)$%
\end{tabular}
\right.  $.

\item[D5] $D(A,B)\vee D(C,E)\geq\left\{
\begin{tabular}
[c]{l}%
$D(A\vee C,B\vee E)$\\
$D(A\wedge C,B\wedge E)$%
\end{tabular}
\right.  $.

\item[D6] $D(A,B)\vee D(A,C)\geq\left\{
\begin{tabular}
[c]{l}%
$D(A,B\vee C)$\\
$D(A,B\wedge C)$%
\end{tabular}
\right.  $.

\item[D7] $D(A,C)\wedge D(B,C)\leq\left\{
\begin{tabular}
[c]{l}%
$D(A\vee B,C)$\\
$D(A\wedge B,C)$%
\end{tabular}
\right.  $.

\item[D8] $D(A,B)\geq\left\{
\begin{tabular}
[c]{l}%
$D(A\vee C,B\vee C)$\\
$D(A\wedge C,B\wedge C)$%
\end{tabular}
\right.  $

\item[D9a] $D(A,A\vee B)=D(B,A\wedge B).$

\item[D9b] $D(A,A\wedge B)=D(B,A\vee B).$

\item[D10] All of the following quantities are equal:
\[%
\begin{array}
[c]{llll}%
(i) & D(A,B)\qquad & (iv) & D(A\wedge B,A\vee B)\qquad\\
(ii) & D(A\wedge B,A)\vee D(A,A\vee B)\qquad & (v) & D(A,A\vee B)\vee
D(B,A\vee B)\qquad\\
(iii) & D(A\wedge B,B)\vee D(B,A\vee B) & (vi) & D(A,A\wedge B)\vee
D(B,A\wedge B)
\end{array}
.
\]
\end{description}
\end{proposition}

\begin{proof}
This follows from the definition of $D(A,B)\doteq S^{\prime}(A,B)$ and
Proposition \ref{pro0403}.$\blacksquare$
\end{proof}

\noindent\textbf{Remark}. Property \textbf{D4 }is related to the concepts of
\emph{betweenness} and \emph{\ convexity}. Consider first an abstract metric
space $(\mathbf{X,}d)$, where $d:\mathbf{X}\times\mathbf{X}\rightarrow
\lbrack0,\infty)\ $ is a scalar distance function. A point $b$ is said to be
(on a \emph{straight line segment}) \emph{between} points $a,c\in\mathbf{X}$
iff $d(a,b)+d(b,c)=d(a,b)$ (this is in direct analogy to the case of a
Euclidean space). Now, in complete analogy, in the autometrized space
$(\mathbf{F},D)$, a ``point'' (actually fuzzy set) $C$ is said to be between
$A,B\in\mathbf{F}$ iff $D(A,C)=D(A,B)\vee D(B,C)$. It can be shown
\cite{Blum99} that these $C$ are exactly the ones which also satisfy $A\wedge
C\leq B\leq A\vee C$. In short, ``points'' which are \emph{order}-between $A$
and $C$ are also \emph{metrically}-between $A$ and $C$. The condition $A\leq
B\leq C$, appearing in Property \textbf{D4}, is a special case of the
condition $A\wedge C\leq B\leq A\vee C$. Convexity can be defined as follows:
a collection of fuzzy sets $\mathbf{W\subseteq F}$ is (metric- or order-)
convex iff for any $A,C\in\mathbf{W}$ we have $D(A,C)$ = $D(A,B)\vee D(B,C)$
$\Rightarrow$ $B\in\mathbf{W}$ (i.e. every point between $A$ and $C$ is
contained in $\mathbf{W}$).

\noindent\textbf{Remark}. Similarly, Property \textbf{D10} is related to
lattice modularity. Take a modular lattice $(\mathbf{X},\leq)$ with a
\emph{positive valuation} $v(.)$ and define a distance $d(.,.)$ by
$d(x,y)=v(x\vee y)-v(x\wedge y)$ (for details see \cite{Birk01}). Then it is
easy to prove $d(x\vee y,x\wedge y)$ = $d(x,y)$, which is analogous to the
equality of (i) and (iv) in \textbf{D10}$.$

\begin{proposition}
For all $A,B,C\in\mathbf{F}$ and all $\Theta\in\mathbf{B}$ we have:

\begin{enumerate}
\item $D(A,B)\leq\Theta\Rightarrow D(A\vee C,B\vee C)\leq\Theta$.

\item $D(A,B)\leq\Theta\Rightarrow D(A\wedge C,B\wedge C)\leq\Theta$.

\item $D(A\wedge C,B\wedge C)=D(A\vee C,B\vee C)=\Theta\Rightarrow
D(A,B)=\Theta.$
\end{enumerate}
\end{proposition}

\begin{proof}
This follows from the definition of $D(A,B)\doteq S^{\prime}(A,B)$ and
Proposition \ref{pro0404}.$\blacksquare$
\end{proof}

\section{Conclusion}

\label{sec05}

We have introduced a novel L-fuzzy valued measure of inclusion $I(.,.),$ and
have established a number of its properties. Some of these properties are
analogous to the ones usually postulated for ``scalar'' fuzzy inclusion
measures. In addition, we find it particularly attractive that $I(.,.)$ turns
out to be a fuzzy order. Furthermore, $I(.,.)$ can be used to define a L-fuzzy
similarity and a L-fuzzy distance between fuzzy sets. Let us conclude by
discussing some future research directions.

\noindent\textbf{L-fuzzy order and lattice. }As already mentioned, $I(.,.)$ is
a L-fuzzy order relationship. In fact this order relationship can be denoted
in an alternative, more suggestive manner. Rather than writing $I(A,B)=\Theta$
(where $\Theta\in\mathbf{B}$) we can also write $A\leq_{\Theta}B$. Now, it is
easy to show that for every fixed value of $\Theta$, $\leq_{\Theta}$ is a
\emph{crisp} \emph{preorder }on $\mathbf{F}$. The following problem arises naturally.

\begin{problem}
Define a L-fuzzy lattice in such a manner that the family $\{(\mathbf{F}%
,\leq_{\Theta})\}_{\Theta\in\mathbf{B}}$ are its \emph{cuts}. Develop the
corresponding L-fuzzy lattice theory.
\end{problem}

\noindent\textbf{Implication. }The relationship between set inclusion and
logical implication is a well known one (as discussed in several parts of this
paper). Our L-fuzzy inclusion measure can be viewed as an \emph{L-fuzzy valued
implication operators}. This connection will be reported elsewhere and
parallels the work of many authors in obtaining inclusion measures from fuzzy
implication operators, in the style of Bandler and Kohout. The L-fuzzy
implication operator satisfies properties analogous to Klir's axioms for the
implication operator \cite{Klir01}. The connection to \emph{conditional
probability} is also worth investigating. A scalar inclusion measure $i(A,B)$
is in many ways analogous to $\Pr(B|A)$ (compare with inclusion measure no.1
in Section \ref{sec0302}); it would be interesting to use the ideas presented
in this paper to define lattice-valued probability measures.

\noindent\textbf{Aggregation. }It is interesting to investigate the existence
of other ``vector'' inclusion measures. In particular, ``scalar'' inclusion
measures and our $I(.,.)$ lie at extreme ends of a spectrum. A\ scalar
inclusion measure aggregates the inclusion information of \emph{all} elements
of a set into a single ``global'' value; our $I(.,.)$ preserves all the
``local'' information about elementwise inclusion. Perhaps inclusion measures
which are halfway between extreme aggregation and extreme localization will
also prove useful. A route to arrive at such inclusion measures could be the
following:\ define a partition $\{U_{1},U_{2},...,U_{K}\}$ of $U$ and then
define on $\mathbf{F}\times\mathbf{F}$ a vector inclusion measure
$\widetilde{I}(A,B)$ $\doteq$ \ $[\widetilde{i}_{1}(A,B)$, $\widetilde{i}%
_{2}(A,B)$, ... , $\widetilde{i}_{K}(A,B)]$ where, for $k=1,2,...,K$,
$\widetilde{i}_{k}(A,B)$ is a scalar inclusion measure which depends only on
values of $A_{u}$, $B_{u}$, for $u\in U_{k}$. In this manner, the details
about elementwise inclusion are not aggregated into a single number; some
degree of local information is preserved.

In this connection, let us mention the approach to \emph{similarity} measures
appearing in \cite{Fonck01}. In this paper, the authors view a similarity
measure as the relation resulting from similarity between degrees of
membership as \emph{local} relations. They relate this approach to implication
operators and discuss transitivity. While in the above paper the local
\emph{preferences} are aggregated (unlike our own approach) we find
interesting the explicit recognition that similarity (and inclusion) is
determined in terms of many local relations. This is exactly the approach we
are taking in this paper, except that we do not perform the aggregation step.

\end{document}